\documentclass[journal=nalefd,manuscript=letter,layout=traditional]{achemso}
\usepackage[T1]{fontenc}

\usepackage{geometry}
\usepackage{setspace}
\usepackage{graphicx}
\usepackage{float}
\newfloat{scheme}{htbp}{los}
\floatname{scheme}{Scheme}
\floatname{chart}{Chart}
\newfloat{graph}{htbp}{loh}
\usepackage{graphicx}
\usepackage{color}
\usepackage{url}
\usepackage{amssymb}
\usepackage{float}
\usepackage{graphicx} 
\usepackage{amsmath}
\usepackage[dvipsnames]{xcolor}
\usepackage{soul} 
\usepackage{multirow}
\usepackage{newtxtext}
\usepackage{booktabs}
\usepackage{nicefrac}
\usepackage{gensymb}
\usepackage{hyperref}
\usepackage{chemformula} %allows use of chemical formula \ch{}
\usepackage[dvipsnames, final]{changes}

\usepackage{tabularx}
\usepackage{array}
\usepackage{placeins}

\newcommand{\BostonCollege}{Department of Physics, Boston College, Chestnut Hill, MA 02467, USA}
\newcommand{\StonyBrook}{Department of Physics and Astronomy, Stony Brook University, Stony Brook, NY 11794, USA}
\newcommand{\AFRL}{Materials and Manufacturing Directorate, Air Force Research Laboratory, Wright-Patterson Air Force Base, OH 45433, USA}
\newcommand{\LNLS}{Brazilian Synchrotron Light Laboratory (LNLS), Brazilian Center for Research in Energy and Materials (CNPEM), 13083-970, Campinas, S\~ao Paulo, Brazil}
\author{Gabriel Natale}
\affiliation{\BostonCollege}
\author{Uma Chirkova}
\affiliation{\BostonCollege}
\author{Fl\'{a}vio Henriques Feres}
\affiliation{\StonyBrook}
\alsoaffiliation{\LNLS}
\author{Ran Jing}
\affiliation{\StonyBrook}
\author{Michael Geiwitz}
\affiliation{\BostonCollege}
\author{Wenyao Liu}
\affiliation{\BostonCollege}
\author{Emma Low}
\affiliation{\BostonCollege}
\author{Josh Leeman}
\affiliation{Department of Chemistry, Princeton University, Princeton, NJ 08544, USA}
\author{Kyung-Mo Kim}
\affiliation{\BostonCollege}
\author{Leslie M. Schoop}
\affiliation{Department of Chemistry, Princeton University, Princeton, NJ 08544, USA}
\author{Mohamed Shehabeldin}
\affiliation{\BostonCollege}
\author{Qiong Ma}
\affiliation{\BostonCollege}
\author{Michael A. Susner}
\affiliation{\AFRL}
\author{Pijush Bhattacharya}
\affiliation{\AFRL}
\alsoaffiliation{Azimuth Corporation, a Core4ce LLC Company, 2079 Presidential Dr. No. 200 Fairborn, OH 45342, USA}
\author{Genda Gu}
\affiliation{Condensed Matter Physics and Materials Science, Brookhaven National Laboratory (BNL), Upton, NY 11973, USA}
\author{Katherine Lee}
\affiliation{Department of Applied Physics and Applied Mathematics, Columbia University, New York, NY 10027, USA}
\author{James Hone}
\affiliation{Department of Mechanical Engineering, Columbia University, New York, NY 10027, USA}
\author{Mengkun Liu}
\affiliation{\StonyBrook}
\author{Kenneth S. Burch}
\affiliation{\BostonCollege}
\email{burchke@bc.edu}

\title{Scalable, Simple, and Versatile Encapsulation of 2D Materials and Devices}



\begin{document}
\newpage

\begin{abstract}

\added{Air-sensitive 2D materials present a fundamental challenge for device integration. Encapsulation is often required to preserve intrinsic properties, yet conventional protection strategies often fail for thicker layers and complicate fabrication. Here, we demonstrate that electron-beam (e-beam) evaporated aluminum oxide (\ch{AlO_x}) serves as both an effective encapsulation layer and a platform for direct device fabrication. Unlike transfer-based approaches, this scalable method is compatible with thicker flakes and full device or wafer coverage. It requires no stacking procedures and enables contacts without post-encapsulation etching. Using rare-earth tritellurides (\ch{RTe_3}, R = La, Er), semimetallic \ch{WTe_2}, and superconducting \ch{FeTe_xSe_{1-x}}, we show that \ch{AlO_x} suppresses oxidation and preserves intrinsic optical and electronic properties. We establish substrate-dependent optimization of encapsulation across a range of flake thicknesses, demonstrate that ultrathin \ch{AlO_x} preserves \ch{WTe_2}'s plasmonic response and maintains superconducting performance in \ch{FeTe_xSe_{1-x}}. Thus we overcome the longstanding tradeoff between encapsulation and straightforward device fabrication in fragile quantum materials.}
\end{abstract}

% \maketitle
\added{Despite the remarkable progress of two-dimensional (2D) materials in electronic, photonic, and quantum technologies, their sensitivity to environmental degradation, fabrication-induced disorder, and surface defects remains a significant obstacle to both fundamental studies and device integration.\cite{Rhodes2019,Hus2017} These challenges are particularly severe in air-sensitive materials well beyond the single layer limit. Specifically, oxidation and exposure to fabrication chemicals and resists can rapidly degrade the electronic and structural properties of exfoliated flakes before measurements can be performed. To preserve pristine surfaces, the 2D materials community has largely adopted encapsulation within hexagonal boron nitride (h-BN), owing to its chemical stability, atomically flat surface, and low defect density.\cite{Li2014} While highly effective, h-BN encapsulation also introduces practical limitations. This process is limited to thin stacks, requires specialized transfer techniques, multiple fabrication steps, and polymer supports which can leave residues and often necessitate additional cleaning procedures.\cite{Frisenda2018,Liu2024} Furthermore, device fabrication is typically performed prior to encapsulation, as post-encapsulation processing requires carefully optimized etching procedures to selectively expose contact regions without damaging the underlying material.\cite{Schulman2018,Tsen2015,Tsen2016} In addition, this approach is not scalable and thus there is a pressing need for encapsulation strategies that simultaneously protect sensitive materials while remaining compatible with scalable device fabrication.}

Alternative approaches based on oxide encapsulation have been explored, particularly using atomic layer deposition (ALD). However, ALD processes often require reactive precursors, surface functionalization, water exposure, or aggressive etching procedures that are incompatible with many fragile low-dimensional materials.\cite{Cano2019,Yang2009,Park2017,Sharma2021} In contrast, electron-beam (e-beam) evaporated aluminum oxide (\ch{AlO_x}) can be deposited at room temperature using a simple and widely accessible process. Importantly, thin \ch{AlO_x} layers are readily removed during standard photolithographic development using tetramethylammonium hydroxide (TMAH)-based developers, enabling wafer scale device fabrication free of polymer residue without dedicated etching steps.\cite{Geiwitz2024,GEIWITZ2026,Kumar2022} This unique property allows the active material to remain protected from photoresist residues, solvents, and ambient contamination throughout fabrication while preserving compatibility with conventional top-contact device architectures. Such a strategy offers an attractive alternative to both transfer-based encapsulation schemes and pre-patterned bottom-contact approaches, which often require complex alignment procedures and can suffer from contamination trapped at buried interfaces or introduce unwanted strain.\cite{Liu2024,Frisenda2018}

Here, we demonstrate that e-beam evaporated \ch{AlO_x} serves as both an effective encapsulation layer and a fabrication platform for air-sensitive quantum materials. Using an integrated glovebox-based workflow, we investigate three representative material systems spanning distinct classes of low-dimensional physics, which we in turn probed with a range of experiments.\cite{Gray2020,Geiwitz2024,GEIWITZ2026} First, we establish the effectiveness of \ch{AlO_x} encapsulation in the rare-earth tritellurides \ch{RTe_3} (R = La, Er), a family of high-mobility quasi-two-dimensional metals that host exotic charge-density-wave phases but are highly susceptible to tellurium oxide (\ch{TeO_x}) formation.\cite{Singh2025,Wang2022,Kopaczek2023,Schoop2020,Chen2019} Next, we demonstrate that ultrathin \ch{AlO_x} coatings preserve the intrinsic Raman and near-field optical response of exfoliated \ch{WTe_2}, enabling high-resolution plasmonic measurements without degradation. Finally, we show that both thin protective layers and thick encapsulation layers are compatible with device fabrication in the iron-based superconductor \ch{FeTe_xSe_{1-x}}, preserving superconducting performance across multiple dopings and lithographic techniques while enabling direct fabrication of top contacts. Together, these results establish e-beam evaporated \ch{AlO_x} as a scalable and versatile route for protecting, characterizing, and integrating fragile quantum materials into functional devices.

\begin{figure*}[t]
  \centering
  \includegraphics[width=1\textwidth]{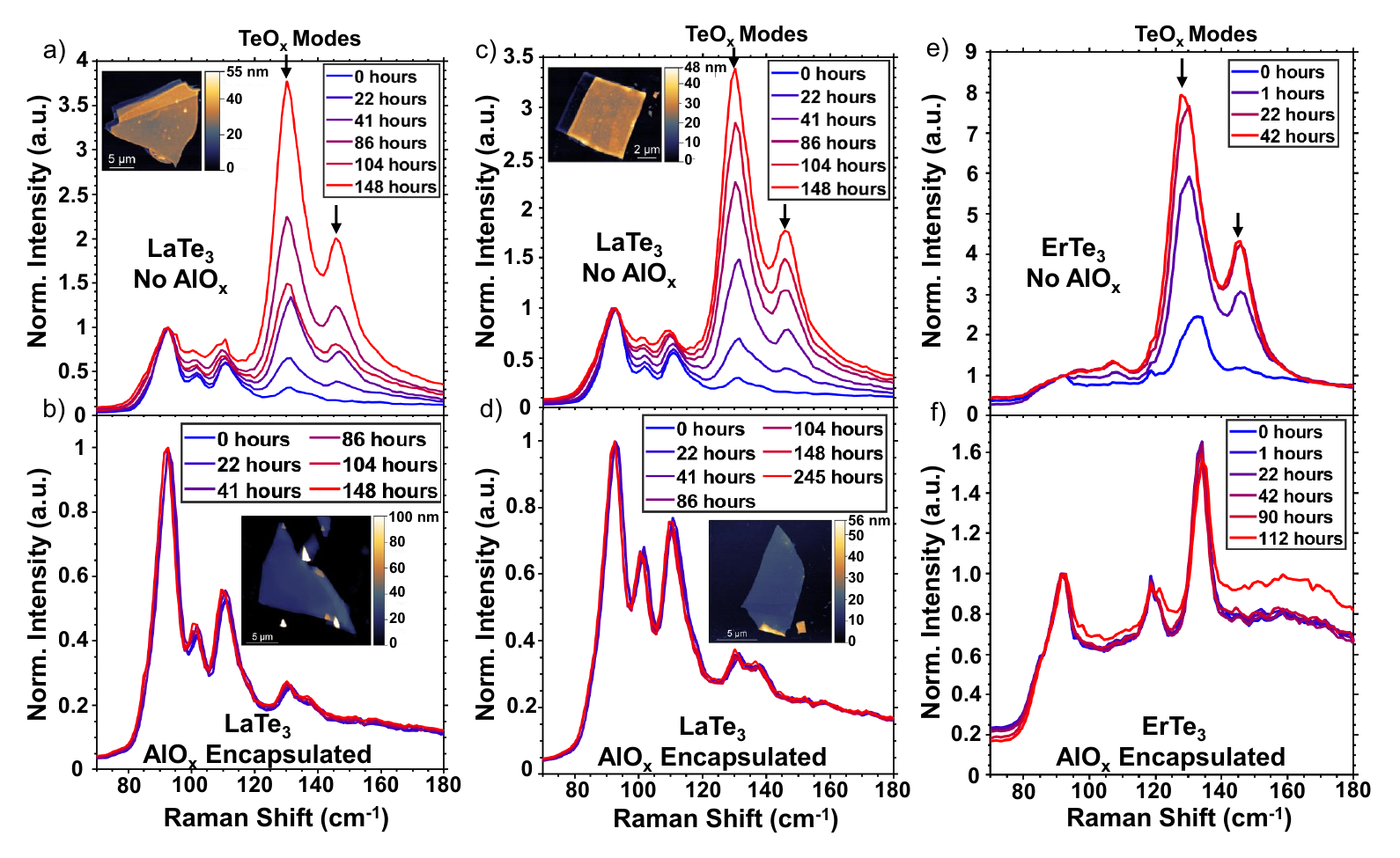}
  \caption{Evolution of thin (\added{$t_{\mathrm{flake}} < 30$ nm}) \ch{LaTe3} and \ch{ErTe3} flakes with exposure to the ambient environment measured via Raman spectra normalized to the lowest energy, intrinsic phonon. The flakes without \ch{AlO_{x}} capping (a,c,e) reveal a rapid and continued enhancement of the \ch{TeO_x} peaks (black arrows), while the \ch{AlO_x} coated flakes (b,d,f) remain unchanged. Insets in a-d display the AFM topography of the flakes, revealing that encapsulated flakes remain uniform, whereas flakes without \ch{AlO_{x}} capping exhibit a rough surface and interlayer delamination.}
  \label{fig:LaTe3AirExposure}
\end{figure*}

The rare-earth tritellurides have garnered widespread interest due to their unique electronic properties and observations of exotic ferroaxial charge density waves.\cite{Singh2025,Schoop2020} However, their stability issues have limited investigations to thick flakes of light rare-earth containing materials, with very little work performed on thin flakes and devices. \cite{Singh2025,Wang2022,Kopaczek2023,Schoop2020,Chen2019} Oxidation is associated with the Te-rich layers and commonly initiates at exposed edges and defects, requiring protection from exfoliation through measurement to avoid degradation. Because \ch{TeO_x} formation produces distinct Raman features, Raman spectroscopy provides a convenient probe of oxidation and encapsulation efficacy.\cite{Kopaczek2023,Gray2020} With this in mind, we began our encapsulation study with \ch{LaTe_{3}}, a moderately air sensitive member of the \ch{RTe_{3}} family where oxidation is clearly observed in the Raman spectra via the appearance of tellurium oxide peaks at 130 and 145 cm$^{-1}$.\cite{Kopaczek2023,Yang2017,2Yang2017,Gray2020} We encapsulated thin flakes (\added{$t_{\mathrm{flake}}< 30$ nm}) exfoliated on \ch{SiO_2} substrates with 130 nm of e-beam evaporated \ch{AlO_x} and compared the properties of these heterostructures to those of uncapped flakes using Raman spectroscopy. We normalized the Raman data using the first mode associated with \ch{LaTe_{3}} (92 cm$^{-1}$) so the relative change in oxide content could be evaluated consistently across samples. As shown in Figure~\ref{fig:LaTe3AirExposure}, we see all of the \ch{LaTe_{3}} bulk Raman modes (92, 101, 110 cm$^{-1}$) as well as the two higher energy peaks (130 and 137 cm$^{-1}$). \cite{Lavagnini2008,Singh2025} After less than 1 day in ambient laboratory conditions (\added{$T = 21\,^\circ\mathrm{C}$}, Rel. Humidity $\sim$35-45\%) the flakes without a capping layer displayed the characteristic \ch{TeO_{x}} modes at 130 cm$^{-1}$ and 145 cm$^{-1}$ (Figure~\ref{fig:LaTe3AirExposure}a,c). With increasing time in ambient conditions the \ch{TeO_{x}} peaks became progressively broader and more intense, dominating over the intrinsic modes of the underlying flake. In stark contrast, the Raman spectra from the encapsulated samples shown in Figure~\ref{fig:LaTe3AirExposure}b,d showed nearly perfect overlap over the course of 6 to 10 days.

\begin{figure*}[!htbp]
  \centering
  \includegraphics[width=1\textwidth]{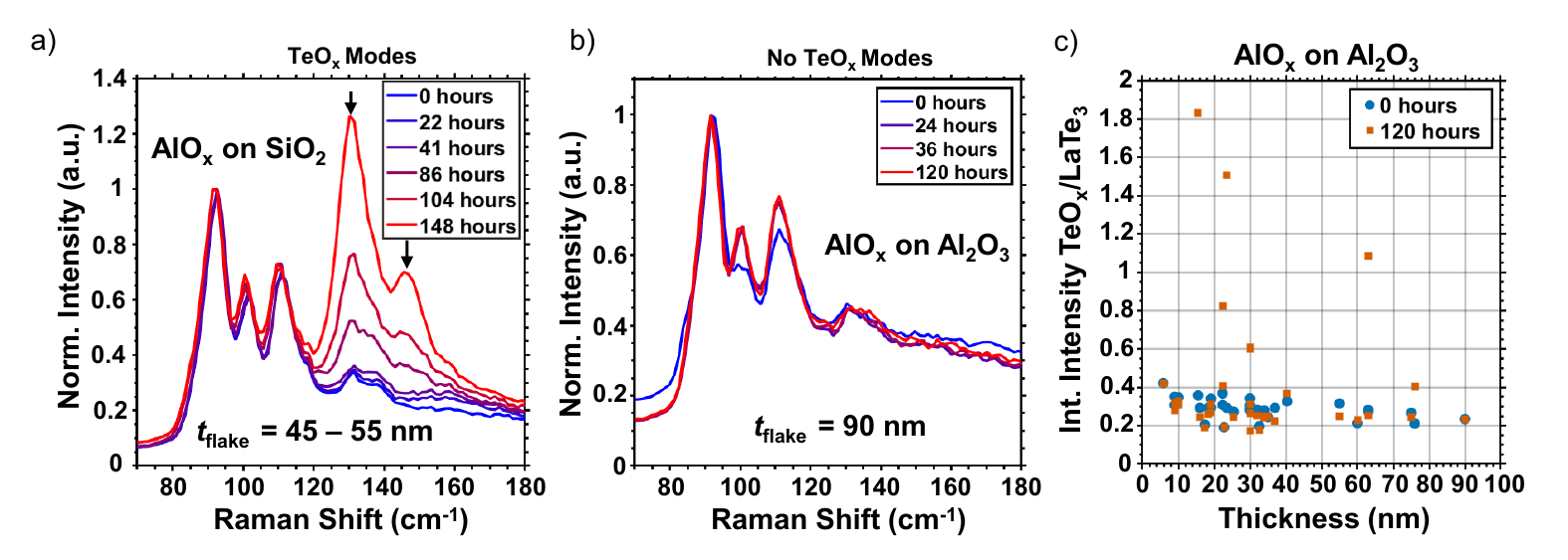}
  \caption{The impact of substrate choice on edge encapsulation quality is demonstrated by comparing the evolution of \ch{LaTe_{3}} Raman spectra normalized to the lowest energy phonon mode for flakes with \added{$t_{\mathrm{flake}}\ge 45$ nm} on a) \ch{SiO_{2}} and b) sapphire (\ch{Al_{2}O_{3}}) substrates. c) Ratio of the integrated intensity of the \ch{TeO_{x}}-related mode (130 cm$^{-1}$) to the lowest energy \ch{LaTe_{3}} mode (92 cm$^{-1}$) vs flake thickness for 37 measurement points across 26 unique flakes.}
  \label{fig:LaTe3Substrate}
\end{figure*}

To test the generality of the \ch{AlO_{x}} encapsulation approach, we examined \ch{ErTe_{3}}, an \ch{RTe_{3}} compound known to exhibit substantially higher air sensitivity than \ch{LaTe_{3}}.\cite{Yumigeta2021} In the absence of encapsulation, \ch{ErTe_{3}} flakes rapidly develop \ch{TeO_{x}}-related Raman features and lose spectral contrast within $\sim$1 hour of ambient exposure (Figure~\ref{fig:LaTe3AirExposure}e). In contrast, as seen in Figure~\ref{fig:LaTe3AirExposure}f \ch{AlO_{x}}-encapsulated \ch{ErTe_{3}} flakes preserve their characteristic phonon modes with no measurable degradation over several days under identical conditions.\cite{Lazarevic2011} The small enhancement of the broad feature at 160 cm$^{-1}$ is likely due to small differences in the collection spot of the laser. These results confirm that the encapsulation strategy remains effective in the extreme limit of air sensitivity within the \ch{RTe_{3}} family.

Following the air-exposure series, we used AFM to assess whether there were any topography changes due to \ch{TeO_{x}} formation. The results are shown in the insets of Figure~\ref{fig:LaTe3AirExposure}a-d, where we observed dramatic effects of oxidation. Specifically, \ch{LaTe3} flakes without \ch{AlO_{x}} capping displayed bubbles and a self-cleaving effect in which layers delaminated. However, these effects were completely absent in the encapsulated flakes. These results suggest the oxidation starts at the edge and proceeds between van der Waals layers, requiring the capping layer to fully cover the edge to ensure protection.

To test this hypothesis, we turned to test flakes of various heights while fixing the encapsulation layer thickness. As seen in Figure~\ref{fig:LaTe3Substrate}a for thicker flakes (\added{$t_{\mathrm{flake}}\ge 45$ nm}) also exfoliated on \ch{SiO_2} substrates, the 130 nm encapsulation layer was not effective. At some point between 41 and 86 hours after exposure to air, the \ch{TeO_{x}}-related modes began to appear, despite initial spectra indicating protection. This suggests the edges were not fully encapsulated and facilitated progressive oxidation through the van der Waals layers. The fact that oxidation begins at the edges was further confirmed by Raman maps on \ch{ErTe_3}, shown in the supplementary information (Figure S4). 

Next, we turned to test if the requirement of extremely thick \ch{AlO_x} relative to the thickness of the underlying flake emerges from its low quality when grown on amorphous \ch{SiO_2}. Specifically, we tested flakes exfoliated on crystalline sapphire (\ch{Al_{2}O_{3}}) substrates and capped them with the same thickness of e-beam deposited \ch{AlO_x}. As seen in Figure~\ref{fig:LaTe3Substrate}b, a 90 nm \ch{LaTe_{3}} flake displayed no evidence of oxidation after 120 hours in air, demonstrating complete edge encapsulation. This was confirmed in a large sampling of flakes of various thicknesses ranging from 5 nm to 90 nm, with many point spectra taken at both the flake center and edge. We illustrate this in Figure~\ref{fig:LaTe3Substrate}c where the ratio of the integrated intensity of the \ch{TeO_{x}} mode to the \ch{LaTe_{3}} mode is plotted as a function of flake thickness immediately following exfoliation/capping and after 120 hours in air. In 32/37 points across 26 unique flakes, the ratios remained largely unchanged, and those 5 outliers can be attributed to specific features of the flakes being probed. For example, flakes with hairline cracks, densely populated regions on the substrate, overlapping flakes, or scratches where the deposited \ch{AlO_{x}} was unable to form a continuous layer. This provides confirmation that by carefully choosing the underlying substrate, we were able to optimize the process such that it could be applied to flakes over a large thickness range. \added{We suspect the difference in encapsulation quality is due to enhanced \ch{AlO_x} nucleation and improved interface morphology when grown on crystalline sapphire substrates as opposed to amorphous \ch{SiO2}. The amorphous substrate increases the likelihood of gaps or voids forming during deposition, making permeation of \ch{O2} and \ch{H2O} more likely with increasing flake thickness.}

\begin{figure*}[!htbp]
  \centering
  \includegraphics[width=1\textwidth]{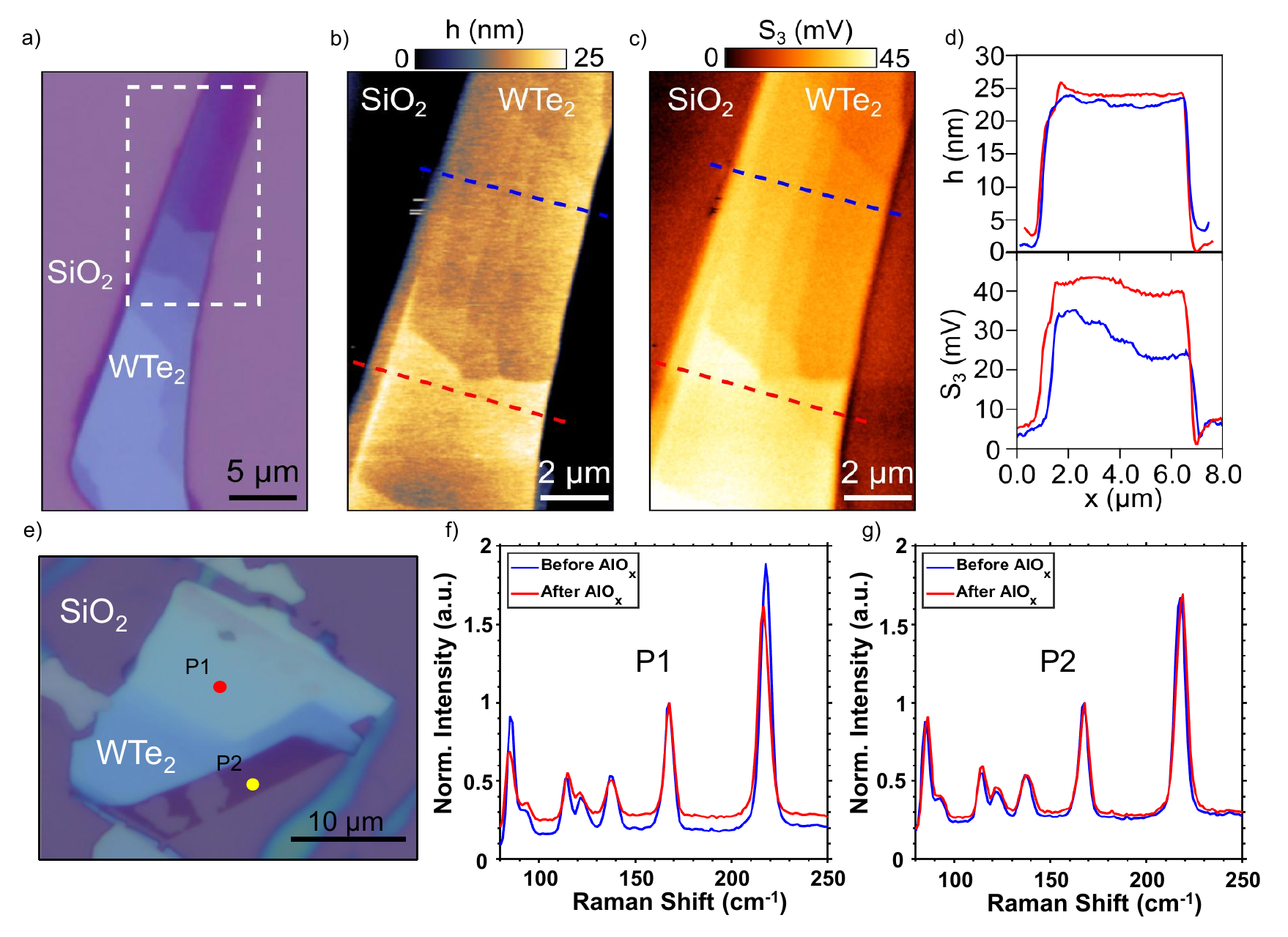}
  \caption{Protection and impact of \ch{AlO_{x}} deposition on \ch{WTe_{2}}. a) Optical microscopy image of the \ch{WTe_{2}} flake on \ch{SiO_{2}} substrate after \ch{AlO_{x}} deposition. b) AFM/topography image and c) 1550 cm$^{-1}$ s-SNOM image performed in the white dashed highlighted area. The blue and red dashed lines indicate the spatial profiles displayed in d) AFM topography (top) and near-field optical image (bottom). e) Optical microscopy image of an exfoliated \ch{WTe_{2}} flake on \ch{SiO_{2}} substrate after \ch{AlO_{x}} deposition with point spectra locations marked in thick (P1) and thin (P2) regions. f) Thick and g) thin region \ch{WTe_{2}} Raman spectra normalized to the peak located at 167 cm$^{-1}$, plotted before (blue) and after (red) \ch{AlO_{x}} deposition.}
  \label{fig:TMD}
\end{figure*}

Having established the effectiveness of \ch{AlO_x} encapsulation for preventing oxidation in \ch{RTe_3}, we next investigated whether ultrathin \ch{AlO_x} layers remain compatible with near-field optical techniques that require nanometer-scale access to the sample surface. We selected \ch{WTe_2} as a stringent test case owing to its pronounced air sensitivity and rich thickness-dependent optical response.\cite{Wang2020} In the plasmonic regime, \ch{WTe_2} exhibits both Drude-like behavior in the THz range and inter-band plasmonic excitations in the far- and mid-infrared spectral ranges.\cite{Jing2021,Xie2023} However, conventional encapsulation layers can attenuate near-field signals and reduce spatial resolution, creating a tradeoff between protection and optical accessibility.\cite{Wang2020}

To evaluate whether e-beam evaporated \ch{AlO_x} can overcome this limitation, we protected mechanically exfoliated \ch{WTe_2} flakes with a 3 nm capping layer and characterized using AFM, Raman spectroscopy, and scattering-type scanning near-field optical microscopy (s-SNOM). An optical image of a representative encapsulated flake is shown in Figure~\ref{fig:TMD}a, while the corresponding AFM topography and near-field optical response are presented in Figure~\ref{fig:TMD}b,c. Figure~\ref{fig:TMD}c presents the 1550 cm$^{-1}$ third-harmonic s-SNOM amplitude image ($\mathrm{S}_3$, see Methods), measured at 20 K. At this frequency, inter-band plasmonic properties of \ch{WTe_2} can be accessed, and these excitations manifest as a pronounced optical contrast across the crystal. The AFM data reveal atomically flat terraces separated by single-layer steps, while the near-field image displays clear thickness-dependent optical contrast between neighboring regions of the flake. This behavior is quantified by the line profiles shown in Figure~\ref{fig:TMD}d, where step heights of approximately 0.6-0.7 nm confirm monolayer differences, and are accompanied by clear changes in the optical signal. The preservation of this thickness-dependent response demonstrates that the ultrathin \ch{AlO_x} layer protects the crystal without obscuring its intrinsic topographic or plasmonic properties, enabling single-layer variations to remain readily distinguishable by s-SNOM.

To further assess whether the deposition process modifies the underlying material, we collected Raman spectra before and after \ch{AlO_x} deposition at the locations indicated in Figure~\ref{fig:TMD}e. As shown in Figure~\ref{fig:TMD}f,g, the spectra remain consistent with previous reports on pristine single crystal \ch{WTe_2} and exhibit no measurable peak shifts, broadening, or changes in relative intensity beyond experimental uncertainty.\cite{Cao2017,Song2016,Buchkov2021} Together, these results demonstrate that ultrathin e-beam evaporated \ch{AlO_x} provides effective protection while preserving both the structural integrity and optical functionality of exfoliated \ch{WTe_2}.

Finally, we explore the use of \ch{AlO_x} as both a protective layer for electrical devices and a platform enabling direct device fabrication. This approach was motivated by our previous work on GFET-based biosensors, in which we employed e-beam evaporated \ch{AlO_x} to shield graphene from photoresist during fabrication and as a passivation layer to isolate the sensing region from the contacts.\cite{Geiwitz2024} During development, the exposed \ch{AlO_x} was removed with only a few additional seconds in developer, while the remainder of the sample remained protected by the 3 nm capping layer.\cite{Geiwitz2024,GEIWITZ2026} This strategy minimizes exposure of the sample surface to photoresist, solvents, and ambient contaminants, and can naturally be extended to air-sensitive materials or systems susceptible to photochemical degradation to enable higher-quality top contacts.

\begin{figure*}[!htbp]
  \centering
  \includegraphics[width=1\textwidth]{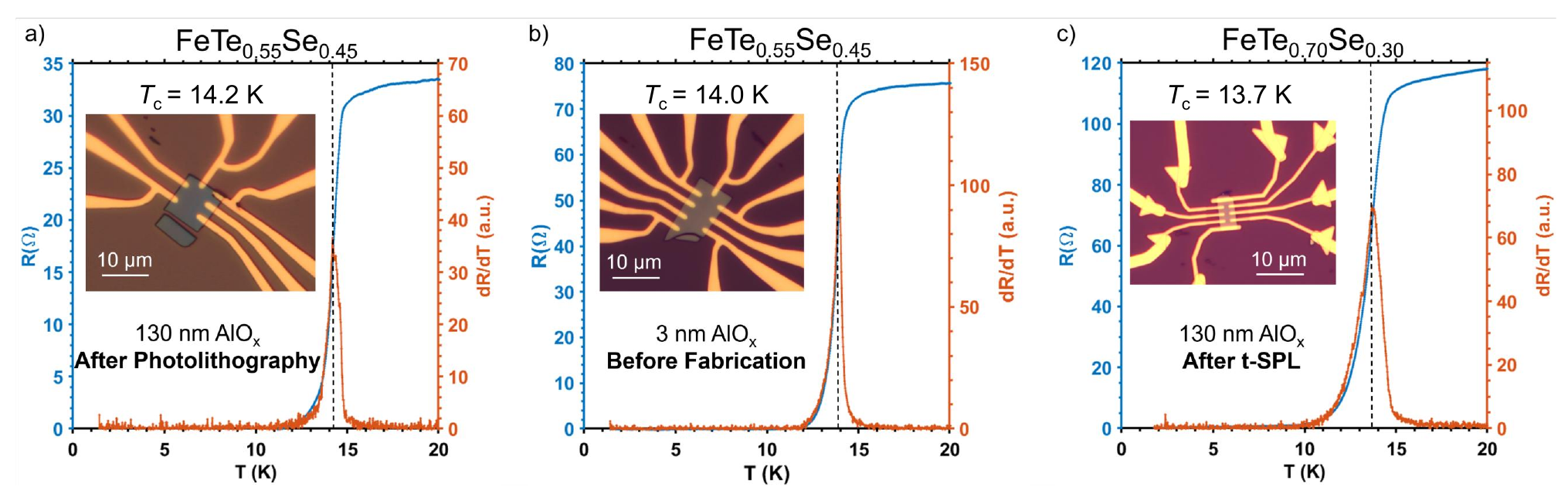}
  \caption{Resistance (blue) and \added{$dR/dT$} (red) vs temperature with device images shown in the inset for a) \ch{FeTe_{0.55}Se_{0.45}} fabricated using photolithography followed by encapsulation b) \ch{FeTe_{0.55}Se_{0.45}} fabricated directly through a 3 nm sacrificial \ch{AlO_x} layer without subsequent thick encapsulation and c) \ch{FeTe_{0.70}Se_{0.30}} fabricated using t-SPL followed by encapsulation. Dashed lines mark the peak position in \added{$dR/dT$}, and thus \added{$T_{\mathrm{c}}$}.}
  \label{fig:FTSComparison}
\end{figure*}

To evaluate the compatibility of this approach with more fragile quantum materials, we investigated the iron-based superconducting family \ch{FeTe_{x}Se_{1-x}}, whose superconducting properties are known to be sensitive to doping, thickness, chemical processing, and air exposure.\cite{Zalic2019,Zhang2018,Zaki2021,Kreisel2020} In addition, clean and stable contacts are particularly important in this system because tunneling measurements have been proposed as a means of probing its potential non-trivial edge states.\cite{gray2019evidence} We mechanically exfoliated optimally doped (highest \added{$T_{\mathrm{c}}$}) \ch{FeTe_{0.55}Se_{0.45}} single crystals onto \ch{SiO_{2}} substrates, and thin flakes (\added{$t_{\mathrm{flake}}< 30$ nm}) were identified by their optical contrast. We coated one sample with a 3 nm \ch{AlO_x} layer prior to resist spin-coating, while we fabricated the remaining devices immediately after exfoliation. We further show that this strategy is compatible with both conventional photolithography (Figure~\ref{fig:FTSComparison}a,b) and thermal scanning probe lithography (t-SPL) (Figure~\ref{fig:FTSComparison}c), the latter enabling sub-200 nm contacts advantageous for spectroscopic tunneling experiments.\cite{Wexler1966,daghero2010probing,Howell2020} Since narrow contacts are particularly susceptible to interface degradation, they stand to benefit significantly from encapsulation strategies that preserve clean interfaces throughout processing. Following fabrication, we further protected selected devices with a 130 nm \ch{AlO_x} encapsulation layer (Figure~\ref{fig:FTSComparison}a,c). Importantly, we performed all fabrication steps without exposing the samples to ambient conditions, ensuring isolation from meaningful levels of oxygen and moisture throughout processing (see supplementary information for details).
      
We wire-bonded samples under ambient conditions and loaded them into a 1.3 K base-temperature transport system within 1 hour for temperature-dependent resistance measurements. All devices exhibited metallic normal-state transport characterized by a monotonically decreasing resistance approaching \added{$T_{\mathrm{c}}$}, followed by a sharp transition to a zero-resistance state, consistent with bulk measurements and the absence of significant disorder.\cite{Kreisel2020,Sacepe2008} As shown in Figure~\ref{fig:FTSComparison}, this behavior was preserved irrespective of doping, lithographic technique, or whether devices were fabricated directly through a 3 nm \ch{AlO_x} protective layer (Figure~\ref{fig:FTSComparison}b), or encapsulated following fabrication (Figure~\ref{fig:FTSComparison}a,c). \added{We identify $T_{\mathrm{c}}$ as the temperature at which $dR/dT$ reaches its maximum, and confirm the validity of this approach by comparing the result obtained for the encapsulated device (Figure~\ref{fig:FTSComparison}a) to another established technique. Namely, by defining $T_{\mathrm{c}}$ as the temperature at which the resistance reaches 50\% of the extrapolated normal-state resistance, determined from a linear fit to $R(T)$ above the superconducting transition (see supplementary information Figure S6). Indeed, we observe excellent agreement between $T_{\mathrm{c}}$ determined using this method ($T_{\mathrm{c}} = 14.21$ K) and the technique which uses the maxima of $dR/dT$ ($T_{\mathrm{c}} = 14.19$ K).} Furthermore, devices with the same composition exhibited consistent \added{$T_{\mathrm{c}}$} values in agreement with those reported in the literature.\cite{Sun_2015,Kreisel2020} The close agreement between directly fabricated and conventionally processed devices demonstrates that we effectively removed the thin \ch{AlO_x} layer in the exposed regions during development, enabling high-quality contacts without compromising superconducting performance. \added{Quantitative comparison of the FWHM of the $dR/dT$ peaks in Figure~\ref{fig:FTSComparison}a,b provides direct confirmation of improved \ch{FeTe_{0.55}Se_{0.45}} device performance when a 3 nm sacrificial mask is employed. Specifically, the FWHM of the directly fabricated device was found to be $\Delta T_{\mathrm{FWHM}} = 0.47$ K while the post-fabrication encapsulated sample displayed $\Delta T_{\mathrm{FWHM}} = 0.77$ K. This demonstrates there is indeed an improvement in quality when using a 3 nm sacrificial layer, and that the transition width is sufficiently narrow in both cases to argue against substantial percolative superconductivity shorting damaged regions of the sample.} Although \ch{FeTe_{x}Se_{1-x}} is relatively tolerant of standard processing chemistry, the enhanced stability observed in the \ch{AlO_x}-protected devices suggests that this approach may be particularly valuable for more fragile 2D materials susceptible to plasma treatments, fabrication-induced contamination, or ambient degradation. Following fabrication, thicker \ch{AlO_x} layers can be deposited to fully encapsulate the completed device, including the contacts and flake edges.

Beyond the specific case of \ch{FeTe_{x}Se_{1-x}}, this strategy offers several practical advantages over established fabrication approaches. Unlike transfer-based h-BN encapsulation, it does not require specialized stacking equipment, polymer-assisted assembly, or selective post-encapsulation etching to expose contact regions. Instead, contact windows are defined during the standard development step, thereby protecting the active material from photoresist residues and liftoff solvents throughout processing. \added{This sacrificial mask technique can be employed with resist stacks of various thickness as its etch rate is much faster than that of most resists and requires just a few additional seconds in developer when a 3 nm layer is employed. However, one must be careful to increase the development time just enough to remove the \ch{AlO_x} in the exposed region to avoid over-etching which can cause the pattern to collapse.} Our \ch{AlO_x} based approach may therefore provide an attractive alternative to pre-patterned bottom contacts, which require precise alignment during transfer and can suffer from incomplete interfacial contact arising from trapped contamination or surface roughness. By enabling direct top-contact fabrication on freshly exposed surfaces using conventional cleanroom tools, e-beam evaporated \ch{AlO_x} simplifies device fabrication while preserving compatibility with subsequent encapsulation. While future work is required to establish its ability to achieve high-quality ohmic contacts in a broader range of quasi-two-dimensional materials, we provide the foundation for such efforts by demonstrating that \ch{AlO_x} is non-destructive to exfoliated \ch{RTe3} flakes, \ch{WTe_{2}} multilayers (as well as 2H-\ch{MoTe_{2}} and \ch{WSe_{2}}, see supplementary information Figure S2) and to \ch{FeTe_{x}Se_{1-x}} devices while enabling direct fabrication of top contacts suitable for transport measurements. \added{Additionally, preliminary Raman and s-SNOM measurements on monolayer, bilayer, and trilayer \ch{WSe2} and \ch{WTe2} demonstrate the promise of this approach for single layer flakes, but further optimization is required (see supplementary information Figure S3 and Figure S5).}

\added{In conclusion, we demonstrate that e-beam evaporated \ch{AlO_x} provides a simple, scalable, and non-destructive platform for the protection and device integration of fragile and thicker quantum materials. Through systematic studies of \ch{RTe_3}, \ch{WTe_2}, and \ch{FeTe_xSe_{1-x}}, we show that \ch{AlO_x} effectively suppresses oxidation, preserves intrinsic optical and electronic properties, and remains compatible with both near-field optical measurements and superconducting device fabrication. Importantly, we establish that thin \ch{AlO_x} layers can serve not only as encapsulation layers but also as sacrificial fabrication masks, enabling direct top-contact fabrication without dedicated post-encapsulation etching steps or exposure of the active material to photoresist residues and processing chemistry.}

\begin{table}[!htbp]
\centering
\caption{Comparison of commonly used encapsulation strategies for fragile
two-dimensional materials. $\checkmark$ = inherent or demonstrated
advantage, $\triangle$ = process- or
material-dependent, and $\times$ = generally
not an advantage of the method.}
\label{tab:encapsulation_comparison}

\renewcommand{\arraystretch}{1.25}
\setlength{\tabcolsep}{5pt}

\begin{tabularx}{\textwidth}{
    >{\raggedright\arraybackslash}X
    >{\centering\arraybackslash}p{2cm}
    >{\centering\arraybackslash}p{2cm}
    >{\centering\arraybackslash}p{2cm}}
\toprule
\textbf{Characteristic} &
\textbf{E-beam \ch{AlO_x}} &
\textbf{ALD \ch{Al2O3}} &
\textbf{h-BN} \\
\midrule

Wafer-scale / large-area compatible
    & $\checkmark$ & $\checkmark$ & $\triangle$ \\

\textbf{Waterless process}
    & $\checkmark$ & $\triangle$\textsuperscript{a} & $\checkmark$\textsuperscript{b} \\

\textbf{Avoids reactive chemical precursors/co-reactants}
    & $\checkmark$ & $\times$ & $\checkmark$ \\

\textbf{Uniform growth without surface treatment}
    & $\checkmark$\textsuperscript{c} & $\triangle$ & $\checkmark$ \\

Room-temperature deposition/assembly possible
    & $\checkmark$ & $\triangle$ & $\checkmark$ \\

\textbf{Efficient deposition of thick ($>$100 nm) encapsulation layers}
    & $\checkmark$ & $\triangle$ & $\times$ \\

Highly conformal coating
    & $\times$ & $\checkmark$ & $\times$ \\

\textbf{No deterministic transfer or alignment step}
    & $\checkmark$ & $\checkmark$ & $\times$ \\

Polymer/stamp free
    & $\checkmark$ & $\checkmark$ & $\triangle$\textsuperscript{d} \\

Ultrathin encapsulation possible
    & $\checkmark$ & $\checkmark$ & $\checkmark$ \\

Optically transparent encapsulation
    & $\checkmark$ & $\checkmark$ & $\checkmark$ \\

\textbf{Contact windows via TMAH-based developer chemistry}
    & $\checkmark$\textsuperscript{e}
    & $\checkmark$\textsuperscript{f}
    & $\times$ \\

Encapsulation of complete device in a single step
    & $\checkmark$ & $\checkmark$ & $\triangle$\textsuperscript{g} \\

\bottomrule
\end{tabularx}

\vspace{0.6em}

\begin{minipage}{\textwidth}
\footnotesize

\textsuperscript{a}
Conventional \ch{Al2O3} ALD commonly employs trimethylaluminum
(TMA) and \ch{H2O}, although alternative water-free chemistries are available.

\textsuperscript{b}
Dry-transfer h-BN encapsulation can avoid exposure of the active material
to water; specific transfer procedures vary.

\textsuperscript{c}
In the present approach, e-beam \ch{AlO_x} is deposited directly onto the material without requiring a nucleation layer, surface functionalization, or chemically reactive precursor sequence. In contrast, uniform ALD nucleation on pristine van der Waals surfaces can be
process-dependent.

\textsuperscript{d}
Conventional h-BN encapsulation typically employs polymer-based stamps and requires precise alignment/transfer, although alternative transfer approaches exist.

\textsuperscript{e}
In the present method, exposed e-beam \ch{AlO_x} is removed during TMAH-based photoresist development, enabling contact-window formation without a separate dry-etch step.

\textsuperscript{f}
ALD \ch{Al2O3} has also been demonstrated to be removable using TMAH-based developer chemistry; the required development/etch time depends on the ALD film properties and processing conditions.

\textsuperscript{g}
Complete h-BN encapsulation typically requires separate top and bottom encapsulation layers, sequential transfer/assembly steps, and can not be deposited across multiple devices in a single step.

\end{minipage}

\end{table}

Unlike transfer-based encapsulation approaches, this strategy does not require specialized stacking equipment, polymer-assisted assembly, or precise alignment to pre-patterned bottom contacts. Instead, it integrates naturally with conventional lithographic workflows while maintaining protection throughout fabrication. The demonstration of stable and scalable encapsulation across a broad range of material classes, measurement techniques, substrate choices, and flake thicknesses highlights the versatility of the approach and suggests a general route toward overcoming the longstanding tradeoff between encapsulation and device fabrication. \added{A direct comparison with alternative techniques is provided in Table~\ref{tab:encapsulation_comparison}.} We anticipate that this platform will facilitate the study and integration of air-sensitive quantum materials into future electronic, photonic, and quantum devices while providing a practical alternative to existing encapsulation technologies.
%%%%%%%%%%%%%%%%%%%%%%%%%%%%%%%%%%%%%%%%%%%%%%%%%%%%%%%%%%%%%%%%%%%%%
\FloatBarrier

\section{Supporting Information}
See supplementary information for details on crystal growth, sample preparation, device fabrication, experimental methods, additional supporting data and discussion, AFM characterization of the deposited \ch{AlO_x} surface, additional \ch{WTe_{2}} s-SNOM data, Raman spectra on 2H-\ch{MoTe_{2}} and \ch{WSe_{2}} flakes before and after \ch{AlO_x} encapsulation, additional \ch{ErTe_{3}} Raman data and spatial Raman mapping, and encapsulated \ch{FeTe_{0.55}Se_{0.45}} resistance vs temperature data and analysis over an extended temperature range (PDF)

\section{Author Information}
\subsection{Corresponding Author}
Kenneth S. Burch - Department of Physics, Boston College, Chestnut Hill, MA 02467, USA

\subsection{Author Contributions}
K.S.B. conceived and supervised the project; G.N., U.C., K.K. and K.S.B. designed and conducted the Raman spectroscopy experiments; Exfoliation and AFM characterization was performed by G.N., U.C., and E.L., with assistance from Q.M., and M.S.; \ch{FeTe_{x}Se_{1-x}} devices were fabricated by G.N. with assistance from M.G. and W.L.; Transport measurements were conducted by G.N. with assistance from W.L.; Cryo s-SNOM and in-situ AFM was performed by F.H.F., R.J., and M.L.; L.M.S. and J.L. provided the \ch{RTe_{3}} crystals; K.L. and J.H. provided the \ch{MoTe_{2}} crystals; P.B. and M.A.S. provided the \ch{FeTe_{0.55}Se_{0.45}} crystals; G.G. provided the \ch{FeTe_{0.70}Se_{0.30}} crystals; G.N. drafted the manuscript with the help of K.S.B.; All authors contributed to the discussion of the manuscript.

\subsection{Notes}
The authors declare no competing financial interest.
The authors would like to thank the Boston College Integrated Sciences Cleanroom for assistance with the work presented in this paper.

\section*{Acknowledgements}
The Nanofrazor used in this research was supported by the NSF MRI Program under the award number 2117711. \ch{MoTe_{2}} crystal synthesis was supported by NSF MRSEC program at Columbia through the Center of Precision-Assembled Quantum Materials (DMR-2011738). Crystal growth and characterization at the Air Force research laboratory was performed under the auspices of the Air Force Office of Scientific Research, LRIR 23RXCOR003 and LRIR 26RXCOR010. F.H.F acknowledges FAPESP Young Investigator process 2019/14019-7. F.H.F acknowledges FAPESP Post-doc project process 2025/00060-0. The work at BNL was supported by the US Department of Energy, office of Basic Energy Sciences, contract no. DOE-SC0012704. The \ch{RTe3} synthesis was supported by an NSF CAREER grant (DMR-2144295) to L.M.S. Support for initial characterization of \ch{RTe3}, Raman and transport experiments was provided by the Air Force Office of Scientific Research under award number FA9550-24-1–011. The device fabrication was supported by the Air Force Office of Scientific Research under award number FA2386-24-1-4071. Q.M. and M.S. acknowledge support from the U.S. Department of Energy, Office of Basic Energy Sciences, under Award No. DE-SC0026332. 

\newpage
\bibliography{Bibliography}

\end{document}

\maketitle
\newpage
\section{Methods} 
To preserve sample quality and prevent oxidation, all device fabrication and characterization from exfoliation through metal deposition and encapsulation was performed in our 'Clean-room in a Glovebox,' which allows us to make complex devices in an inert argon environment without exposing samples to ambient conditions.\cite{Gray2020} Thermal scanning probe lithography (t-SPL) using the Nanofrazor was performed in a separate glovebox located in the same building. Samples were spin coated and sealed in an argon filled vacuum flange before transferring them between gloveboxes. 

\subsection{\ch{RTe_{3}} Growth and Exfoliation}
High-quality \ch{RTe_{3}} (R=La, Gd, Ho or Er) single crystals were grown in an excess of tellurium (Te) by a self-flux technique. Te (metal basis $>$99.999$\%$, Sigma-Aldrich) was first purified to remove oxygen contamination and then mixed with a rare earth ($>$99.9$\%$, Sigma-Aldrich) in a ratio of 97:3. The mixture was sealed in an evacuated quartz ampoule and heated to 900 $^\circ$C over 12 hours and then slowly cooled down to 550 $^\circ$C at a rate of 2 $^\circ$C/h. The crystals were separated from the flux by centrifugation at 550 $^\circ$C. 

\ch{RTe_{3}} single crystals were mechanically exfoliated inside an inert argon glovebox with 0.15 ppm \ch{O_{2}} content using Scotch tape to prevent oxidation of the air-sensitive material. To reduce flake thickness, the tape was repeatedly folded and pressed against the crystal. The tape, now bearing thin \ch{RTe_{3}} flakes, was then pressed onto pre-cleaned \ch{Si/SiO_{2}} or sapphire (\ch{Al_{2}O_{3}}) substrates. For flakes exfoliated onto \ch{Si/SiO_{2}} substrates, optical microscopy images were acquired under standard white-light illumination (Zeiss Axioscope 7). Green channel intensity values of the substrate background and the candidate flake were extracted using ImageJ, and the optical contrast $C$ was calculated as $C$ = ($I_{\mathrm{background}}$ – $I_{\mathrm{flake}}$)/$I_{\mathrm{background}}$. For flakes exfoliated onto sapphire substrates, thickness was assessed using transmission-mode optical microscopy in the green channel, where thinner flakes appear brighter due to increased light transmission. Color contrast values were extracted from transmission images using ImageJ and used to determine thickness, which was subsequently verified by atomic force microscopy (AFM) measurements on selected flakes.

\subsection{\ch{Fe(Te,Se)} Growth and Exfoliation} 
\ch{FeTe_{0.70}Se_{0.30}} single crystals with high quality were grown in the BNL group by an unidirectional solidification method.\cite{wen2011interplay} 

The AFRL group synthesized \ch{FeTe_{0.55}Se_{0.45}} from high purity elements using the following process. First, we mixed Fe granules (Alfa Aesar, 99.99$\%$), Se shot (Alfa Aesar, 99.999$\%$), and Te ingot (Alfa Aesar, 99.9999+$\%$) in the desired stoichiometric ratio and melted in a round-bottom quartz ampoule (sealed under 1/3 atm UHP Ar) at a temperature of $\sim$1080$^\circ$C for $\sim$24 h followed by a slow cool (20$^\circ$C/hr) to room temperature. The homogenized material was extracted from the quartz, sealed in a Bridgman quartz ampoule (14 mm ID, 19 mm OD, 20 cm long) under the same conditions, and lowered at a rate of $\sim$1 cm/day through a Bridgman furnace set-up where the hot side was held at 1050 $^\circ$C  and the cold side was $\sim$800 $^\circ$C. After careful extraction from the ampoule, we were able to cleave a natural surface and check composition with a Hitachi TM 4000 Plus mated with an Oxford Aztec One Xplore30 Compact EDS system. Structure was verified with a Malvern PanAlytical Empyrian X-ray Diffractometer using Cu K-alpha radiation.

The bulk Fe(Te,Se) crystals were mechanically exfoliated inside an inert argon glovebox with 0.15 ppm \ch{O_{2}} onto pre-cleaned \ch{Si/SiO_{2}} substrates using the Scotch tape method and thin flakes ($t_{\mathrm{flake}}<30$ nm) were identified by their contrast using optical microscopy (Zeiss Axioscope 7), and for some flakes later confirmed via AFM.

\subsection{2H-\ch{MoTe_{2}}, \ch{WSe_{2}}, and \ch{WTe_{2}} Growth and Exfoliation}

Molybdenum ditelluride (\ch{MoTe_{2}}) crystals were synthesized by the Columbia group using a two-step self-flux method with a metal to chalcogen molar ratio of 1:5. Mo powder of 99.997$\%$ purity was loaded into a quartz ampule with Te ingot of 99.999+$\%$ purity and sealed under high vacuum ($\sim$1e$^{-5}$ Torr). This ampule was heated to 1120 $^\circ$C for 24 h, held for 3 days, then cooled at a rate of 1 $^\circ$C/h to 750 $^\circ$C, followed by an increased cooling rate of 5 $^\circ$C/h to 600 $^\circ$C, before rapidly cooling to room temperature. The ampule contents were then transferred to a new ampule and annealed across a temperature gradient for 48 h with the hot end at 470 $^\circ$C and the cold end at room temperature. 

High quality \ch{WSe_{2}} crystals were obtained from HQ Graphene.

High quality \ch{WTe_{2}} crystals were obtained from 2D Semiconductors.

The 2H-\ch{MoTe_{2}}, \ch{WSe_{2}}, and \ch{WTe_{2}} crystals were also mechanically exfoliated inside an inert argon glovebox with 0.15 ppm \ch{O_{2}} onto pre-cleaned \ch{SiO_{2}} substrates using the Scotch tape method and thin flakes ($t_{\mathrm{flake}}<30$ nm) were identified by their contrast using optical microscopy.

\subsection{\ch{Fe(Te,Se)} Device Fabrication}
\paragraph{\ch{FeTe_{0.70}Se_{0.30}}:}
The exfoliated \ch{FeTe_{0.70}Se_{0.30}} samples were spin-coated in our fabrication glovebox with a bi-layer resist stack consisting of a 120 nm PMMA/MA underlayer (Allresist AR-P 617.03) and a 25 nm PPA layer (Allresist Phoenix 81, 2wt\% PPA in Anisole). They were then sealed in a flange and transferred to the glovebox containing the Nanofrazor ($<$0.1 ppm \ch{O_{2}}). The nanofrazor's in-situ AFM was employed for precise alignment of the contact pattern to the flake edges, and was simultaneously used to confirm the flake thickness was in the thin ($t_{\mathrm{flake}}<30$ nm) regime. After patterning the $\sim$250 nm wide contacts using t-SPL, the sample was transferred back to our 'Cleanroom in a Glovebox' and developed using a 5\% DI water in IPA solution. They were then loaded into a thermal deposition tool with in situ \ch{Ar/O_{2}} plasma, and pumped down to pressure $\sim1\times10^{-7}$ torr. A 60 s 60 W Ar plasma de-scumming was then applied immediately before the deposition of 5 nm \ch{Cr}/40 nm \ch{Au}. After soaking the sample in 80 $^\circ$C PG Remover, lift-off was performed, it was rinsed in IPA, and blown dry with Argon. An additional round of fabrication using direct-write photolithography (Heidelberg $\mu$PG101) was employed to pattern the leads and bonding pads. A 100 nm lift off resist layer (LOR1A) was first spin coated, followed by a 500 nm photoresist layer (S1805). The sample was then loaded into the direct-write system, the leads and pads carefully aligned to the existing contacts, and the resist exposed. It was developed in a TMAH-based developer (MF-321/319) and the same plasma, deposition, and liftoff procedure as described above was applied, although a slightly thicker Au layer (55 nm) was deposited to ensure continuity between the stitched leads and the contacts.

\paragraph{\ch{FeTe_{0.55}Se_{0.45}}:}
The \ch{FeTe_{0.55}Se_{0.45}} devices were fabricated using exclusively photolithography following the same spin-coating, writing, development, plasma, and metallization procedure as above to create $\sim$1~$\mu$m wide 5 nm \ch{Cr}/45 nm \ch{Au} contacts along the flake edges. In the case of the device pre-coated with a 3 nm protective \ch{AlO_{x}} layer, the development time was increased by $\sim$2 s to ensure it was adequately removed in the exposed area. 

Finally, devices targeted for encapsulation were placed on the e-beam evaporator's sample plate, and the bonding pads masked off using kapton tape to ensure the gold surface would remain exposed for wirebonding, while the rest of the device was covered with 130 nm \ch{AlO_{x}}. Details of the evaporation process can be found in the following section.  

\subsection{Aluminum Oxide Evaporation}
Electron beam evaporation was performed using precursor aluminum oxide crystals in an Angstrom Engineering system with in-situ plasma. Fresh aluminum oxide crystals were used for each deposition to improve the stability of the deposition rate and the resulting film quality. A slow deposition rate of (0.5 A/s for 130 nm, and 0.3 A/s for 3 nm) was used to achieve a uniform deposition with as few voids as possible. Since substrate heating and \ch{O_2} flow were not used to improve crystallinity and replenish oxygen that may have been stripped from the \ch{AlO_{x}} crystals during the evaporation, the encapsulation layer is likely oxygen deficient (hence our use of \ch{AlO_x} when referring to our encapsulation layer).\cite{MAITI2010214,Hoffman1971} While \ch{O_{2}} flow is not feasible given this technique is intended for samples susceptible to oxidation, we note that substrate heating during the deposition would likely improve the quality of the encapsulation layer and lead to fewer voids, so it is recommended to do this in samples that can tolerate high temperatures ($\sim$ 300 $^\circ$C).

\section{Raman Measurements}
Raman measurements were performed using a Witec Raman system in an inert Argon glovebox with 0.15 ppm \ch{O_{2}} with a 532 nm excitation wavelength. The 1800 gr/mm grating was employed to achieve high resolution spectra focused on peaks found at low energies. The laser power and acquisition times varied between samples, but were generally 150-350 $\mu$W, and 180-240 s. Given the strong Raman response of \ch{RTe_{3}} samples, one must be careful to use lower power so that the detector does not saturate. Raman mapping was performed using the same measurement setup, and a 20 x 20 $\mu m^{2}$ area was scanned with 60 points per line and 60 lines per image, achieving a 0.33 $\mu$m$^{2}$ pixel size.  

\section{AFM Measurements}
AFM measurements were performed using a Park Systems NX-AFM in non-contact (tapping) mode. The raw data was processed (flattened) and analyzed in Gwyddion to accurately determine step heights, flake thicknesses, surface height distributions, and RMS roughness (R$_{q}$) for a given scanning area. 

\section{Transport Measurements}
Transport measurements were performed in a liquid helium sub-cooled variable temperature insert (VTI) with a base temperature of 1.3 K (Cryo Industries of America). Before loading the samples, they were mounted to a 24-pin DIP chip carrier and within 1 hour wire bonded in ambient conditions and loaded into the VTI. They were then cooled to base temperature, and resistance vs temperature curves were taken in a 4-point configuration to avoid contact resistance contributions. A Keithley 6221 precision current source was used to apply a 25 $\mu$A DC bias, and the resulting voltage was measured using an AG34401a digital multimeter. Resistance vs temperature data was collected while warming the sample at a constant rate of 0.5-1 K/min using a custom Labview interface.

\section{Cryo s-SNOM Measurements} 
The scattering-type scanning near-field optical microscope (s-SNOM) employs a metallic atomic force microscope (AFM) tip as an optical nanoprobe to probe the local optical near field. During operation, the AFM tip oscillates in tapping mode at its mechanical resonance frequency. Infrared illumination from a quantum cascade laser (QCL) is focused onto the tip–sample junction using a parabolic mirror, generating a highly confined electromagnetic field at the tip apex through the lightning-rod effect. This interaction induces an effective local polarizability in the tip–sample system. The backscattered signal S, collected by the same parabolic mirror and detected with a mercury–cadmium–telluride (MCT) detector, contains contributions from both the genuine near-field interaction and undesired far-field background scattering. To isolate the near-field component, the scattered signal is demodulated at higher harmonics n$\Omega$ of the tip oscillation frequency using a lock-in detection scheme combined with pseudo-heterodyne detection. Since the near-field contribution dominates at higher harmonics, authentic near-field signals are obtained for n$\ge$2. In this work, the measurements were performed using the third harmonic (n=3).

Infrared nano-imaging experiments under low temperature were carried out using a home-built s-SNOM integrated into a closed-cycle cryostat system (OptiCool, Quantum Design), allowing measurements in the temperature range from approximately 10 to 350 K. A tunable mid-infrared QCL source (Hedgehog, DRS Daylight Solutions) was focused onto the AFM tip via a parabolic mirror. The scattered light carries information about both the local material response. Near-field detection was performed using self-homodyne interferometric schemes together with higher-harmonic lock-in demodulation to suppress far-field contributions and enhance near-field sensitivity. The experiments employed an Akiyama-type scanning probe with a resonance frequency of approximately $\Omega_\mathrm{tip}\approx 65$ kHz, combined with Attocube nanopositioning stages.

\section{Supplementary Aluminum Oxide Characterization}
To confirm the deposited thickness and characterize the surface, a 130 nm layer was deposited on a blank \ch{SiO_{2}} substrate as described above. This was followed by photolithography to define a large window where the \ch{AlO_{x}} was subsequently removed using 65:35 diluted TRANSETCH-N (Transene) for 14 min at 80 $^\circ$C, and then rinsed thoroughly with DI water. The remaining photoresist was then removed using 80 $^\circ$C PG Remover, and the sample was rinsed in IPA and blown dry with Argon. \par
\begin{figure}[!htbp]
  \centering
  \includegraphics[width=1\textwidth]{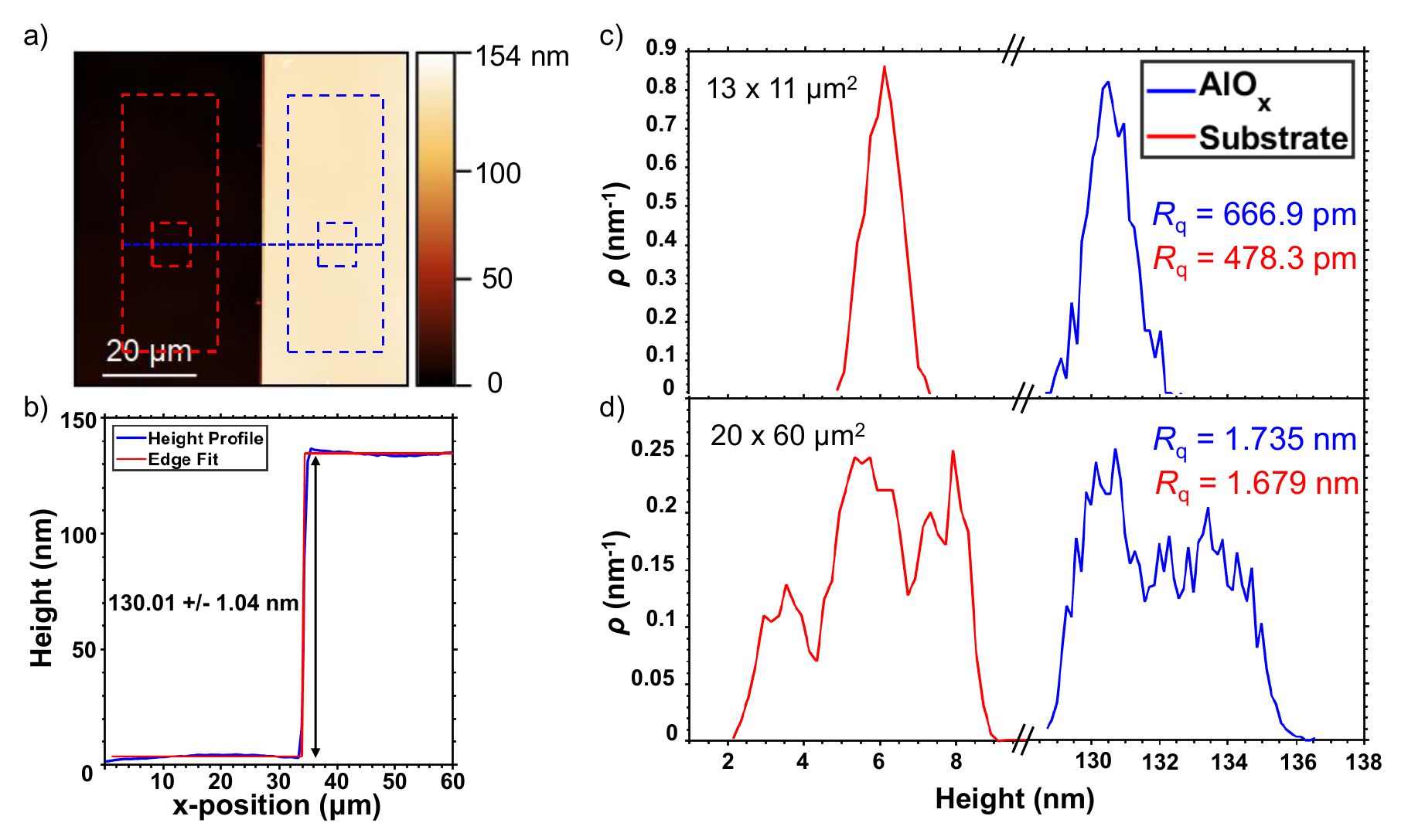}
  \caption{a) \ch{AlO_{x}}/Substrate topography. The solid blue line indicates the position of the line profile in b) and blue and red dashed boxes indicate the areas used to determine the height distribution and roughness of the \ch{AlO_{x}} surface and substrate in c,d). b) line profile taken across the step from substrate to \ch{AlO_{x}} and the resulting step edge fit confirming the film thickness and uncertainty. c) Height distribution over small areas (small dashed red and blue boxes) on the substrate and \ch{AlO_x} surface, inset displays the RMS roughness parameters. d) Height distribution over large areas (large dashed red and blue boxes) on the substrate and \ch{AlO_x} surface, inset displays the RMS roughness parameters.}
  \label{fig:AlOxSurface}
\end{figure}

 AFM was then performed to directly probe the step height between the substrate and the \ch{AlO_{x}} and confirmed the thickness of the deposited layer was indeed 130 nm (See Figure~\ref{fig:AlOxSurface}a,b). In Figure ~\ref{fig:AlOxSurface}c,d the root mean square (RMS) surface roughness ($R_\mathrm{q}$) of both the \ch{AlO_{x}} (blue dashed boxes) and substrate (red dashed boxes) were evaluated by first taking the height distributions of small (13 x 11 $\mu$m$^{2}$) and large (20 x 60 $\mu$m$^{2}$) sample areas. While 1-2 nm of variation is visible when looking at the large area height distribution, this is expected given the wet-oxide \ch{SiO_{2}} layer on any given wafer is 285 nm +/- 5 $\%$ at any point on its surface. Both the substrate and \ch{AlO_{x}} have very similar RMS roughness on both large and small scales (Figure~\ref{fig:AlOxSurface}c,d inset), suggesting the roughness/non-uniformity of the \ch{AlO_{x}} is established by the substrate and is not a result of poor deposition conditions. We note AFM is not sufficient to directly identify voids in the film, so another imaging technique like (Cross-sectional SEM or TEM) must be employed to get a more detailed view of growth defects on the atomic scale.

\FloatBarrier

\section{Supplementary 2H-\ch{MoTe_{2}}, \ch{WSe_{2}}, and \ch{ErTe3} Raman Results}
\begin{figure}[!htbp]
  \centering
  \includegraphics[width=1\textwidth]{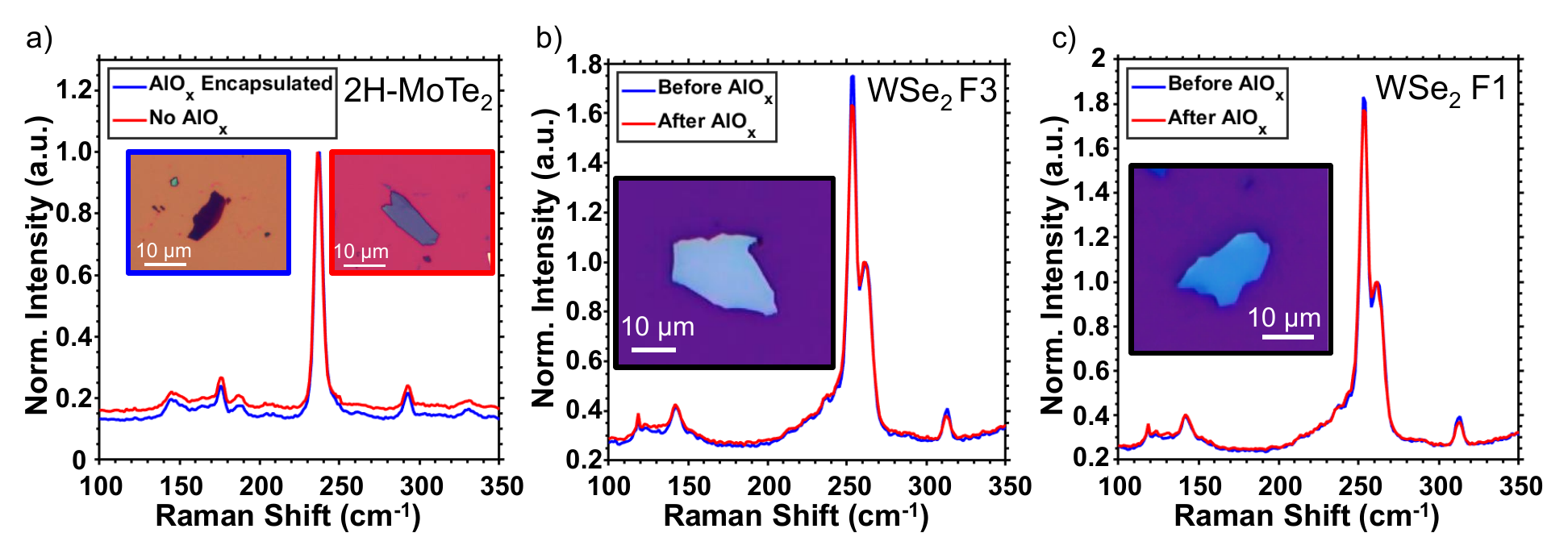}
  \caption{a) Raman spectra taken on 130 nm \ch{AlO_{x}} encapsulated (blue) and unencapsulated (red) 2H-\ch{MoTe_{2}} flakes of similar thickness, normalized to the most prominent peak at 236 cm$^{-1}$. Inset optical microscopy images of the flakes. b,c) Raman spectra taken on 2 \ch{WSe_{2}} flakes before (blue) and after (red) depositing  130 nm \ch{AlO_{x}}, normalized to the peak at 261 cm$^{-1}$. Inset optical microscopy images of the flakes.}
  \label{fig:TMDRaman}
\end{figure}

\begin{figure}[!htbp]
  \centering
  \includegraphics[width=1.0\textwidth]{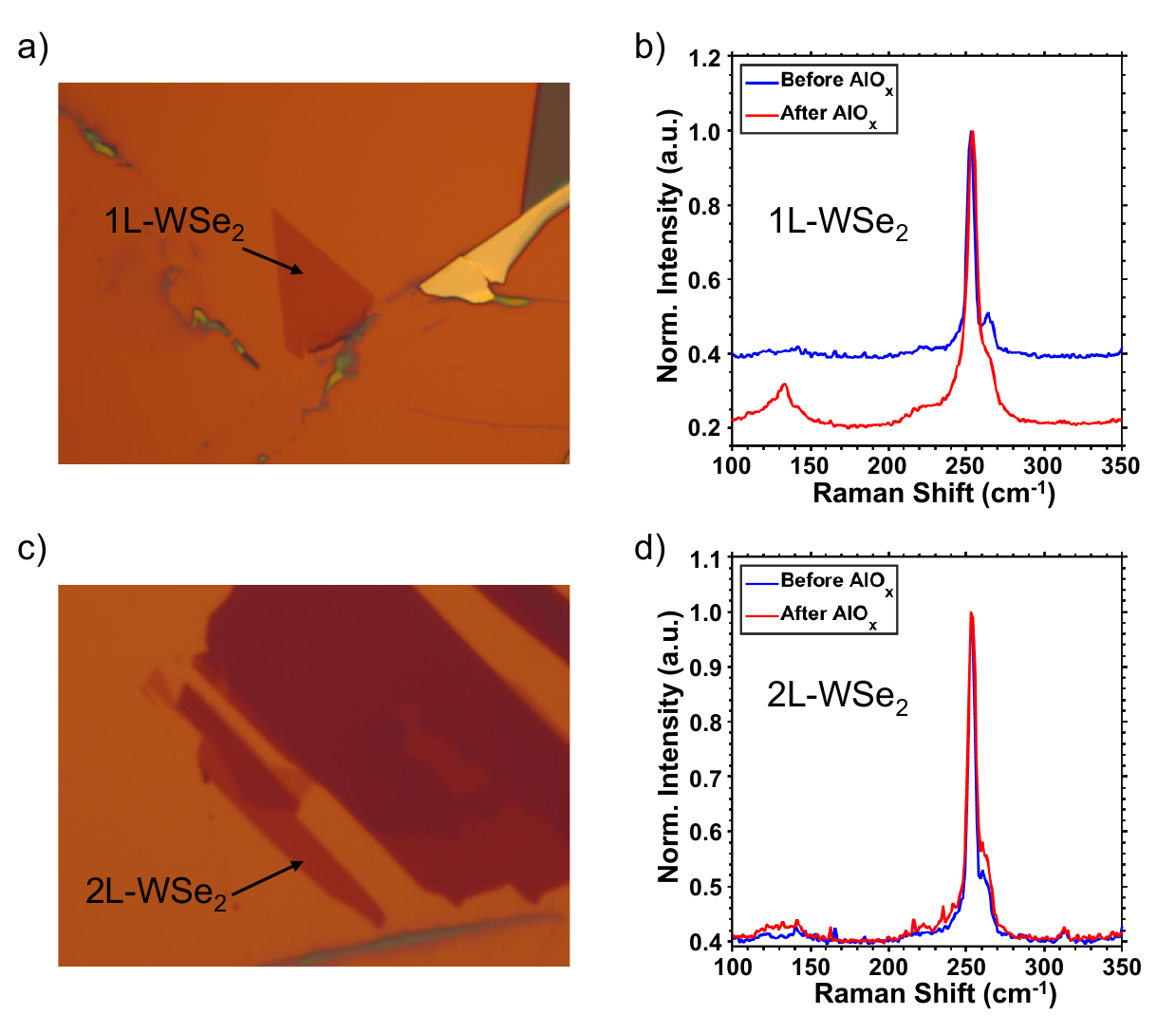}
 \caption{a) Optical microscope image of a monolayer \ch{WSe_{2}} flake. b) Raman spectra normalized to the 253 cm$^{-1}$ peak taken on a monolayer flake before and after encapsulation with 20 nm \ch{AlO_x}. c) Optical microscope image of a bilayer \ch{WSe_{2}} flake. d) Raman spectra normalized to the 253 cm$^{-1}$ peak taken on a bilayer flake before and after encapsulation with 20 nm \ch{AlO_x}.}
  \label{fig:Wse2Thin}
\end{figure}

To obtain further confirmation of the non-destructive nature of this technique to quasi-2D TMD's, Raman spectroscopy was employed on 2H-\ch{MoTe_{2}} and \ch{WSe_{2}}. Flakes were exfoliated onto \ch{Si/SiO_{2}} substrates using the scotch tape method, and two 2H-\ch{MoTe_{2}} flakes of similar thickness were selected for comparison (Figure~\ref{fig:TMDRaman}a insets). One was encapsulated with 130 nm \ch{AlO_{x}} (blue), while the other was left unencapsulated (red) and Raman spectra were taken on both samples. Near perfect overlap was observed in the Raman response when normalized to the highest intensity peak at 236 cm$^{-1}$ (Figure~\ref{fig:TMDRaman}a), again suggesting the process was non-destructive. We note that the spectra also display the expected peaks observed in previous work on pristine thin flakes of this material (170 cm$^{-1}$, 235 cm$^{-1}$, and $\sim$ 288 cm$^{-1}$).\cite{Rani2018} Similarly, as seen in Figure~\ref{fig:TMDRaman}c,d) Raman spectra were taken on two \ch{WSe_{2}} flakes before (blue) and after (red) depositing  130 nm \ch{AlO_{x}}, normalized to the peak at 261 cm$^{-1}$. The excellent overlap between the spectra in both cases demonstrate the process was non-destructive and agrees with previous work on pristine exfoliated flakes.\cite{Terrones2014} Additional Raman measurements on ultrathin exfoliated \ch{WSe2} flakes demonstrate preliminary evidence that this process can likely be optimized for use with monolayer systems in the future, and is compatible with bilayer flakes without any further optimization. In Figure~\ref{fig:Wse2Thin}a,b we show an optical microscope image of a monolayer flake and the Raman spectra normalized to the most prominent \ch{WSe2} peak at $\sim$ 253 cm$^{-1}$ taken before and after e-beam encapsulation with 20 nm \ch{AlO_x}. We found that while there was some impact on the Raman spectrum following the deposition, the main feature corresponding to \ch{WSe2} at $\sim$ 253 cm$^{-1}$ was preserved, suggesting there was not catastrophic damage to the underlying lattice. In Figure~\ref{fig:Wse2Thin}c,d we show an optical microscope image of a bilayer flake and the corresponding normalized Raman spectra taken before and after e-beam encapsulation. The spectra display excellent agreement, and the characteristic Raman modes of the bilayer \ch{WSe2} remain observable following the deposition, with no substantial changes in their spectral positions or linewidths.

\begin{figure}[!htbp]
  \centering
  \includegraphics[width=1\textwidth]{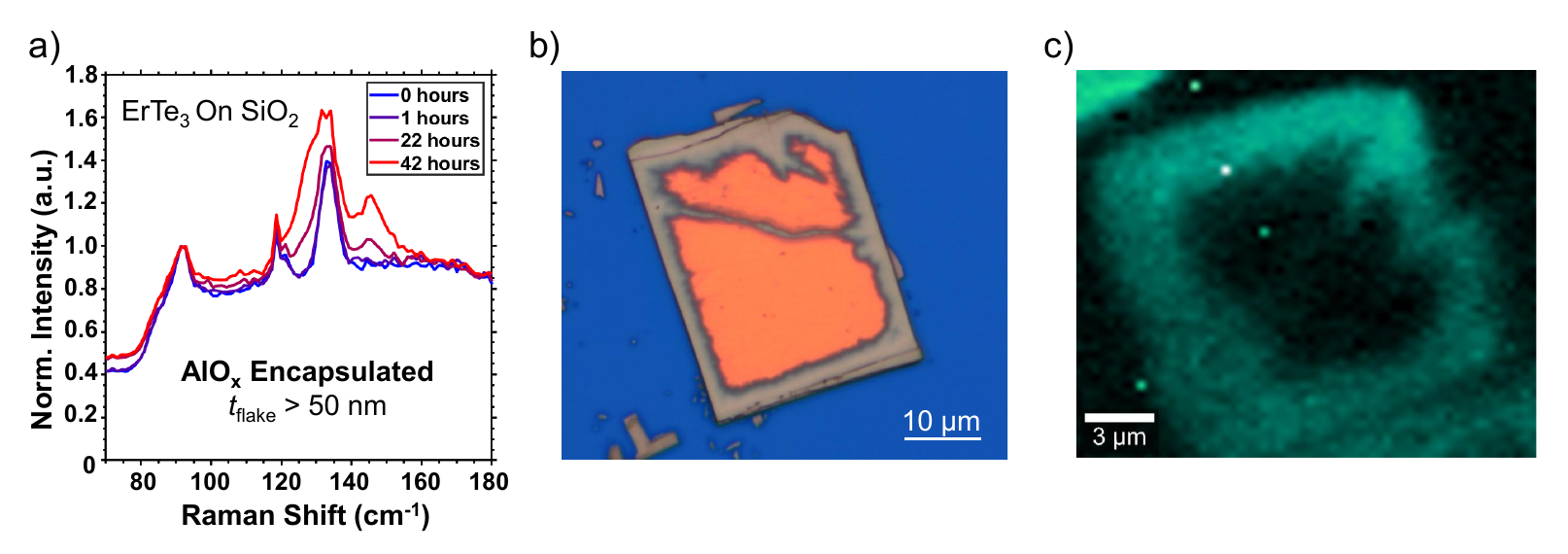}
  \caption{a) Raman spectra normalized to the 92 cm$^{-1}$ phonon mode taken on a 130 nm \ch{AlO_{x}} encapsulated thick ($t_{\mathrm{flake}}\sim 90$ nm) \ch{ErTe_{3}} flake on \ch{SiO_{2}} substrate with increasing exposure to the ambient environment. Demonstrates lack of complete edge encapsulation for thick flakes on \ch{SiO_{2}} substrates. b) Optical microscopy image of a 130 nm \ch{AlO_{x}} encapsulated thick flake, displaying clear signs of edge oxidation. c) Spatial Raman map of the intensity of the \ch{TeO_x}-related mode at 130 cm$^{-1}$ showing enhanced intensity due to oxidation at flake edges.}
  \label{fig:ErTe3SI}
\end{figure}
\FloatBarrier

\section{Supplementary Cryo s-SNOM Results}
\begin{figure}[!htbp]
  \centering
  \includegraphics[width=0.8\textwidth]{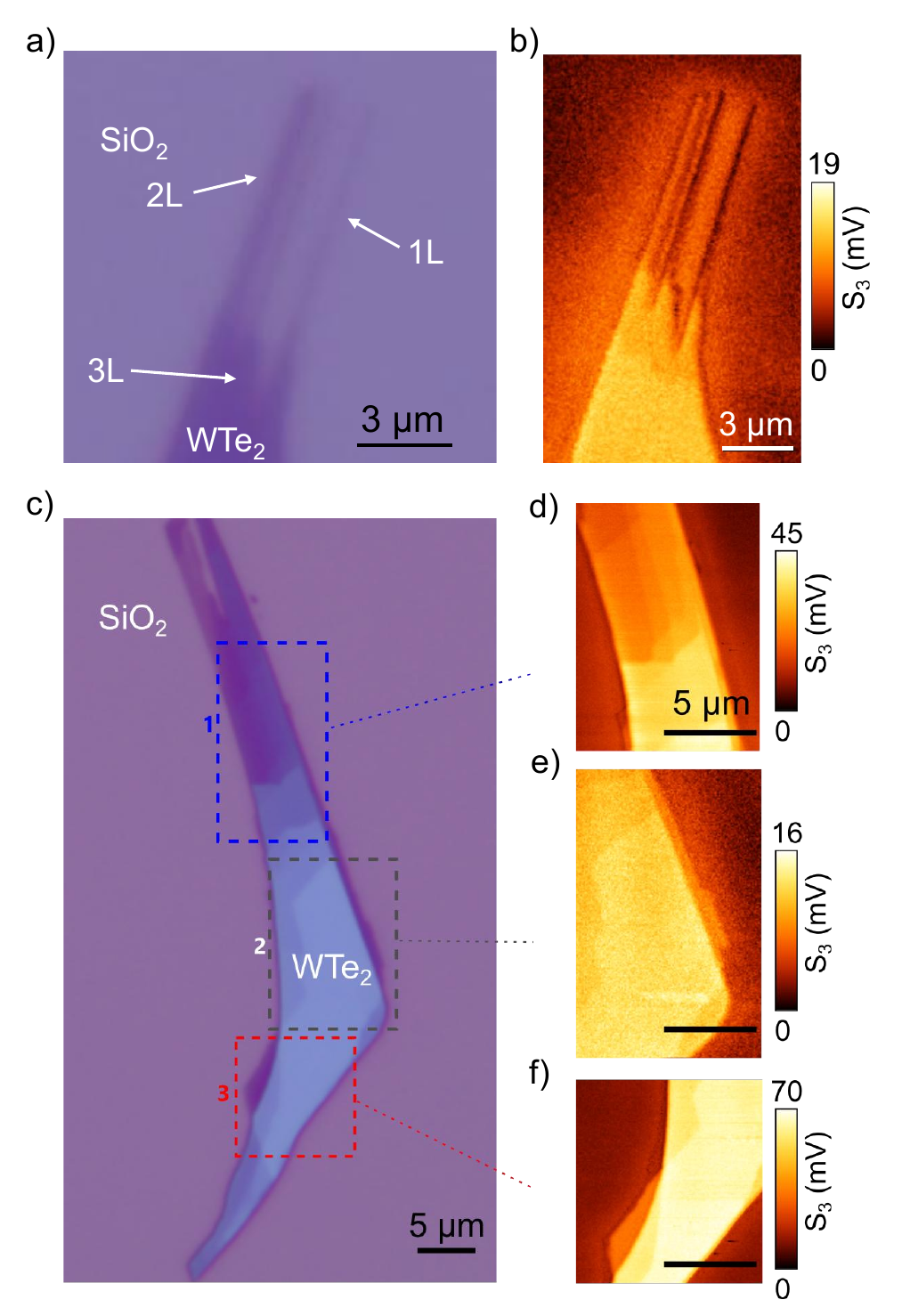}
  \caption{a) Optical microscope image of a 3 nm \ch{AlO_{x}} capped thin (mono to few-layer) \ch{WTe_{2}} flake. b) 1550 cm$^{-1}$ s-SNOM image, showing the contrast between the thin \ch{WTe{2}} crystal and the \ch{SiO_{2}} substrate. Distinct signal observed in all regions except the ultra-thin (most likely monolayer) region. c) Full optical microscope image of the 3 nm \ch{AlO_{x}} capped \ch{WTe_{2}} flake appearing in the main text figure. d, e, f) 1550 cm$^{-1}$ s-SNOM images taken in the boxed regions labeled 1, 2, and 3 respectively, showing the contrast between the \ch{WTe{2}} crystal and the \ch{SiO_{2}} substrate.}
  \label{fig:sSNOM2}
\end{figure}

Here, we present additional s-SNOM measurements performed on a thinner \ch{WTe2} crystal. Figure~\ref{fig:sSNOM2}a,b show the optical microscope image and the corresponding near-field amplitude map, respectively. Even for thinner flakes, a measurable near-field contrast can still be observed. However, in the monolayer limit, there is no clear evidence that the encapsulation layer is sufficient to fully prevent oxidation.
Furthermore, Figure~\ref{fig:sSNOM2}c–f display near-field images acquired in different regions of the flake presented in the main text. In all investigated areas, both the plasmonic features and the surface flatness remain well preserved, indicating the robustness of the sample quality across the crystal. Further work is required to optimize this process for monolayer samples.

\FloatBarrier

\section{Supplementary Transport Results}

\begin{figure}[!htbp]
  \centering
  \includegraphics[width=1\textwidth]{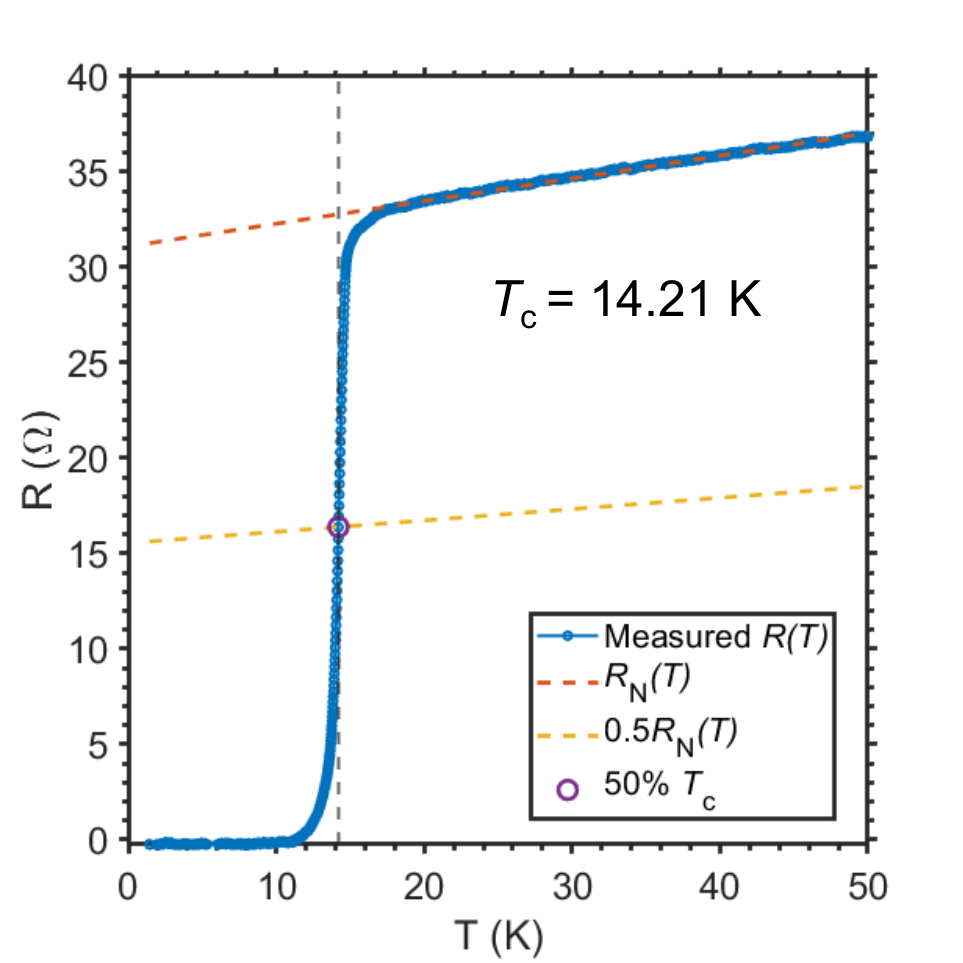}
 \caption{Resistance vs temperature data for the encapsulated device in main text Figure 4a over an extended temperature range, up to 50 K. $T_{\mathrm{c}} = $ 14.21 K is identified by finding the temperature at which the resistance reaches 50\% of the normal-state resistance, determined from a linear fit to $R(T)$ above the superconducting transition.}
  \label{fig:TcFromRN}
\end{figure}

In Figure~\ref{fig:TcFromRN}, we present the resistance vs temperature data obtained for the encapsulated device in main text Figure 4a over an extended temperature range, and identify the superconducting transition temperature, $T_{\mathrm{c}}$, using a 50\% normal-state resistance criterion for comparison to the result obtained from the maximum of $dR/dT$. The normal-state resistance was obtained by performing a linear fit to the $R(T)$ data over a temperature range above the onset of the superconducting transition (16 K to 50 K) and extrapolating the fit through the transition region. $T_{\mathrm{c}}$ was then defined as the temperature at which the measured resistance reached 50\% of the extrapolated normal-state resistance, i.e., $R(T_{\mathrm{c}})$ = 0.5$R_{\mathrm{N}}(T_{\mathrm{c}})$. The crossing temperature was determined by linear interpolation between the nearest measured data points on either side of the 50\% normal-state resistance value.

\clearpage
\bibliography{Bibliography}